\documentstyle[preprint,tighten,aps,amstex,graphicx,times,floats]{revtex}

\begin{document}

\thispagestyle{empty}

\marginparwidth 1.cm
\setlength{\hoffset}{-1cm}
\newcommand{\mpar}[1]{{\marginpar{\hbadness10000%
                      \sloppy\hfuzz10pt\boldmath\bf\footnotesize#1}}%
                      \typeout{marginpar: #1}\ignorespaces}
\def\mda{\mpar{\hfil$\downarrow$\hfil}\ignorespaces}
\def\mua{\mpar{\hfil$\uparrow$\hfil}\ignorespaces}
\def\mla{\marginpar[\boldmath\hfil$\rightarrow$\hfil]%
                   {\boldmath\hfil$\leftarrow $\hfil}%
                    \typeout{marginpar: $\leftrightarrow$}\ignorespaces}

\renewcommand{\abstractname}{Abstract}
\renewcommand{\figurename}{Figure}
\renewcommand{\refname}{Bibliography}

\newcommand{\eg}{{\it e.g.}\;}
\newcommand{\ie}{{\it i.e.}\;}
\newcommand{\etal}{{\it et al.}\;}
\newcommand{\ibid}{{\it ibid.}\;}

\newcommand{\mx}{M_{\rm SUSY}}
\newcommand{\pt}{p_{\rm T}}
\newcommand{\et}{E_{\rm T}}
\newcommand{\del}{\varepsilon}
\newcommand{\sla}[1]{/\!\!\!#1}
\providecommand{\SP}{\scriptscriptstyle}

\newcommand{\nni}{\tilde{\chi}_i^0}
\newcommand{\nnj}{\tilde{\chi}_j^0}
\newcommand{\nne}{\tilde{\chi}_1^0}
\newcommand{\nnz}{\tilde{\chi}_2^0}
\newcommand{\nnd}{\tilde{\chi}_3^0}
\newcommand{\nnv}{\tilde{\chi}_4^0}
\newcommand{\cpi}{\tilde{\chi}_i^+}
\newcommand{\cpj}{\tilde{\chi}_j^+}
\newcommand{\cpe}{\tilde{\chi}_1^+}
\newcommand{\cpz}{\tilde{\chi}_2^+}
\newcommand{\cmi}{\tilde{\chi}_i^-}
\newcommand{\cmj}{\tilde{\chi}_j^-}
\newcommand{\cme}{\tilde{\chi}_1^-}
\newcommand{\cmz}{\tilde{\chi}_2^-}
\newcommand{\cpme}{\tilde{\chi}_1^\pm}
\newcommand{\cpmz}{\tilde{\chi}_2^\pm}
\newcommand{\cpmi}{\tilde{\chi}_i^\pm}

\providecommand{\nni}{\tilde{\chi}_i^0}
\providecommand{\nnj}{\tilde{\chi}_j^0}
\providecommand{\nne}{\tilde{\chi}_1^0}
\providecommand{\nnz}{\tilde{\chi}_2^0}
\providecommand{\nnd}{\tilde{\chi}_3^0}
\providecommand{\nnv}{\tilde{\chi}_4^0}
\providecommand{\cpi}{\tilde{\chi}_i^+}
\providecommand{\cpj}{\tilde{\chi}_j^+}
\providecommand{\cpe}{\tilde{\chi}_1^+}
\providecommand{\cpz}{\tilde{\chi}_2^+}
\providecommand{\cmi}{\tilde{\chi}_i^-}
\providecommand{\cmj}{\tilde{\chi}_j^-}
\providecommand{\cme}{\tilde{\chi}_1^-}
\providecommand{\cmz}{\tilde{\chi}_2^-}

\providecommand{\mni}{m_{\tilde{\chi}_{\SP i}}}
\providecommand{\mnj}{m_{\tilde{\chi}_{\SP j}}}

\providecommand{\mni}{m_{\tilde{\chi}_{\SP i}}}
\providecommand{\mnj}{m_{\tilde{\chi}_{\SP j}}}
\providecommand{\mne}{m_{\tilde{\chi}_{\SP 1}^{\SP 0}}}
\providecommand{\mnz}{m_{\tilde{\chi}_{\SP 2}^{\SP 0}}}
\providecommand{\mnd}{m_{\tilde{\chi}_{\SP 3}^{\SP 0}}}
\providecommand{\mnv}{m_{\tilde{\chi}_{\SP 4}^{\SP 0}}}
\providecommand{\mce}{m_{\tilde{\chi}_{\SP 1}^{\SP +}}}
\providecommand{\mcz}{m_{\tilde{\chi}_{\SP 2}^{\SP +}}}

\providecommand{\msel}{m_{\tilde{e}_{\SP L}}}
\providecommand{\mser}{m_{\tilde{e}_{\SP R}}}
\providecommand{\msne}{m_{\tilde{\nu}_{\SP e}}}

\newcommand{\epc}[3]{${\rm Eur. Phys. J.}$ {\bf C#1} (19#2) #3}
\newcommand{\zpc}[3]{${\rm Z. Phys.}$ {\bf C#1} (19#2) #3}
\newcommand{\npb}[3]{${\rm Nucl. Phys.}$ {\bf B#1} (19#2)~#3}
\newcommand{\plb}[3]{${\rm Phys. Lett.}$ {\bf B#1} (19#2) #3}
\renewcommand{\prd}[3]{${\rm Phys. Rev.}$ {\bf D#1} (19#2) #3}
\renewcommand{\prl}[3]{${\rm Phys. Rev. Lett.}$ {\bf #1} (19#2) #3}
\newcommand{\prep}[3]{${\rm Phys. Rep.}$ {\bf #1} (19#2) #3}
\newcommand{\fp}[3]{${\rm Fortschr. Phys.}$ {\bf #1} (19#2) #3}
\newcommand{\nc}[3]{${\rm Nuovo Cimento}$ {\bf #1} (19#2) #3}
\newcommand{\ijmp}[3]{${\rm Int. J. Mod. Phys.}$ {\bf #1} (19#2) #3}
\renewcommand{\jcp}[3]{${\rm J. Comp. Phys.}$ {\bf #1} (19#2) #3}
\newcommand{\ptp}[3]{${\rm Prog. Theo. Phys.}$ {\bf #1} (19#2) #3}
\newcommand{\sjnp}[3]{${\rm Sov. J. Nucl. Phys.}$ {\bf #1} (19#2) #3}
\newcommand{\cpc}[3]{${\rm Comp. Phys. Commun.}$ {\bf #1} (19#2) #3}
\newcommand{\mpl}[3]{${\rm Mod. Phys. Lett.}$ {\bf #1} (19#2) #3}
\newcommand{\cmp}[3]{${\rm Commun. Math. Phys.}$ {\bf #1} (19#2) #3}
\newcommand{\jmp}[3]{${\rm J. Math. Phys.}$ {\bf #1} (19#2) #3}
\newcommand{\nim}[3]{${\rm Nucl. Instr. Meth.}$ {\bf #1} (19#2) #3}
\newcommand{\prev}[3]{${\rm Phys. Rev.}$ {\bf #1} (19#2) #3}
\newcommand{\el}[3]{${\rm Europhysics Letters}$ {\bf #1} (19#2) #3}
\renewcommand{\ap}[3]{${\rm Ann. of~Phys.}$ {\bf #1} (19#2) #3}
\newcommand{\jhep}[3]{${\rm JHEP}$ {\bf #1} (19#2) #3}
\newcommand{\jetp}[3]{${\rm JETP}$ {\bf #1} (19#2) #3}
\newcommand{\jetpl}[3]{${\rm JETP Lett.}$ {\bf #1} (19#2) #3}
\newcommand{\acpp}[3]{${\rm Acta Physica Polonica}$ {\bf #1} (19#2) #3}
\newcommand{\science}[3]{${\rm Science}$ {\bf #1} (19#2) #3}
\newcommand{\vj}[4]{${\rm #1~}$ {\bf #2} (19#3) #4}
\newcommand{\ej}[3]{${\bf #1}$ (19#2) #3}
\newcommand{\vjs}[2]{${\rm #1~}$ {\bf #2}}
\newcommand{\het}[1]{${\tt hep\!-\!th/}$ {#1}}
\newcommand{\hep}[1]{${\tt hep\!-\!ph/}$ {#1}}
\newcommand{\hex}[1]{${\tt hep\!-\!ex/}$ {#1}}
\newcommand{\desy}[1]{${\rm DESY-}${#1}}
\newcommand{\cern}[2]{${\rm CERN-TH}${#1}/{#2}}

\preprint{
\font\fortssbx=cmssbx10 scaled \magstep2
\hbox to \hsize{
\hskip.5in \raise.1in\hbox{\fortssbx University of Wisconsin - Madison}
\hfill\vtop{\hbox{\bf MADPH-99-1128}
            \hbox{December 1999}} }
}

\title{ 
Measuring CP Violating Phases at a Future Linear Collider 
} 

\author{
V.~Barger, T.~Han, T.~Li, and T.~Plehn
} 

\address{ 
Department of Physics, University of Wisconsin, Madison, WI 53706 
} 

\maketitle 

\begin{abstract}
  At a future Linear Collider one will be able to determine the masses
  of charginos and neutralinos and their pair production cross
  sections to high accuracies. We show how systematically including
  the cross sections in the analysis improves the measurement of the
  underlying mass parameters, including potential CP violating phases.
  In addition, we investigate how experimental statistical errors will
  affect the determination of these parameters. We present a first
  estimate on the lower limit of observable small phases and on the
  accuracy in determining large phases.
\end{abstract} 


\section{Introduction}\label{sec:one}

In the Minimal Supersymmetric Standard Model~\cite{msugra} the gauge
boson and Higgs sectors are mapped to a set of neutral and charged
fermions which mix to form neutralino and chargino mass eigenstates.
The $\tilde{\chi} \tilde{\chi}$ cross sections are parameterized by
the gaugino masses $M_1,M_2$, the higgsino mass parameter $\mu$, and
the ratio of vacuum expectation values $\tan \beta$. In CP
non-conserving Minimal Supergravity models, the parameter $\mu$ can be
complex, and in the most general unconstrained Minimal Supersymmetric
Standard Model the gaugino masses $M_1,M_2$ will also include phases;
after rotating the wino fields, we can without a loss of generality
choose $\phi_2=0$, and are left with two additional phase parameters
in the chargino/neutralino sector:
\begin{equation}
\mu \rightarrow |\mu| \; e^{i \phi_\mu} \qquad \qquad
M_1 \rightarrow |M_1| \; e^{i \phi_1}
\end{equation}

A future Linear Collider~\cite{lincol} will enable the very successful
LEP $e^+e^-$ precision analyses of the Standard Model to be extended
so as to expose any underlying model. From the set of masses, total
cross sections~\cite{cxn} and distributions/asymmetries~\cite{old},
one can unambiguously determine the mass parameters for CP conserving
models.  Yet it has not been quantified how well in a complete
analysis one can measure small~\cite{small,toby} or large~\cite{kane}
complex phases, and how this can improve their analytical
determination from the physical masses~\cite{zerwas,kneur}.
Experimental bounds on electric dipole moments severely restrict the
existence of phases which do not rely on large cancellations, allowing
only small variations due to non-gaugino MSSM parameters, like the
trilinear couplings $A_i$~\cite{small,toby}.  Phases of the order
${\cal O}(\pi/10)$ or even smaller, like ${\cal O}(\pi/100)$, hardly
influence na\"{\i}ve CP conserving observables like masses and total
cross sections and will be difficult to determine at a Linear
Collider. We will give a first estimate of when statistical errors
render small phases unobservable and how well large phases can be
measured, taking into account realistic experimental uncertainties.

\section{Analysis of Masses and Cross Sections}\label{sec:two}

To estimate the effect of CP violating phases on masses and total
cross sections, we examine five scenarios.  Three of them are derived
from a set of mSUGRA parameters taking $m_0=100$~GeV and
$m_{1/2}=200$~GeV~\cite{lincol}, adding small phases $\phi_1$ and
$\phi_\mu$. Since the phase $\phi_1=0.01 \pi$ in scenario (2) is very
small, this parameter set is similar to an mSUGRA scenario. The large
phase scenario~\cite{kane} avoids constraints from electromagnetic
dipole moments~\cite{small,toby} through cancellations.\footnote{The
  very small value of $\tan\beta$ is neither a generic feature of the
  model~\cite{michal} nor of our analysis.  However, we prefer to
  quote the exact values of Ref.~\cite{kane}.}  In the wino lightest
supersymmetric particle model~\cite{randall}, additional phases are
attached to the largest entries $M_1,\mu$ in the neutralino/chargino
mass matrices.  Simply adding small phases to CP conserving models
will reveal how well we will be able to distinguish CP conserving and
CP violating models. Only in the large phase scenario (4) are the
phases a basic feature of the model:

\begin{center} \begin{tabular}{ll|cccc|cc|c}
   && \; $M_1$[GeV] & \; $M_2$[GeV] & \; $\mu$[GeV] & \; $\tan \beta$ \;& 
      \; $\phi_1/\pi$ & \; $\phi_\mu/\pi$ \; & \; $\msel; \mser; \msne$ \\[2mm] \hline 
    (1) \;  & modified mSUGRA \;\; 
  &  82.6   & 164.6   & 310.6   &  4   & 0.1   & 0.1 & 180; 132; 166 \\ 
    (2)     & modified mSUGRA
  &  82.6   & 164.6   & 310.6   &  4   & 0.01  & 0.1 & 180; 132; 166 \\ 
    (3)     & modified mSUGRA
  &  82.6   & 164.6   & 285.0   & 30   & 0.1   & 0.1 & 180; 132; 166 \\ 
    (4) & large phases
  &  75.0   &  85.0   & 450.0   &1.2   & 0.5   & 0.8 & 195; 225; 185 \\ 
    (5) & modified wino LSP
  & 396.0   & 120.0   & 250.0   & 50   & 0.1   & 0.1 & 200; 200; 200 \\ 
\end{tabular} \end{center} \medskip

Mismeasuring masses and total cross sections within their experimental
uncertainties is the limiting feature expected to spoil the
determination or even observation of phases, because the numerical
effect, especially of small phases, is generically small. For each of
the five scenarios we calculate the set of chargino/neutralino masses
and the pair production cross sections for all possible final states.
We regard this sample as a set of experimental observables,
mismeasured with a given Gaussian uncertainty.  Using either a global
fit or the inversion algorithm described in the Appendix, we then
extract again the parameters $M_1,M_2,\mu,\tan\beta$ and
$\phi_1,\phi_\mu$ from the smeared set of observables.\smallskip

\begin{figure}[tbh] 
\begin{center}
\includegraphics[width=9.0cm]{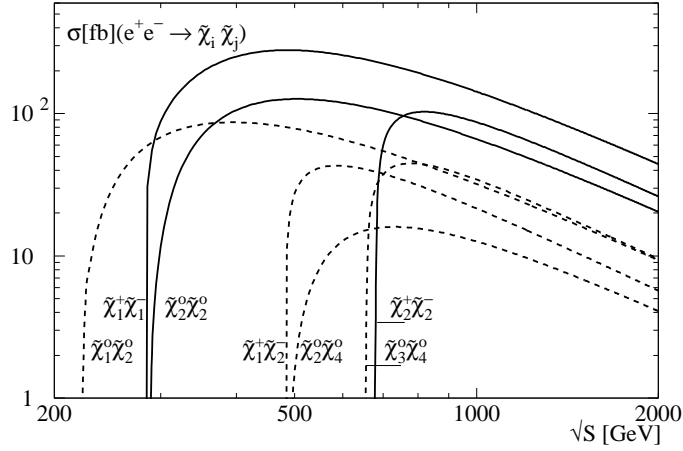} \vspace*{0.0cm}
\caption[]{\label{fg:cxn} 
The largest cross sections for the  
mSUGRA type parameters (1) and 
$\phi_1=0.1 \pi$,
$\phi_\mu=0.1 \pi$ as a function of the collider
energy. 
The solid lines indicate cases where CP violating phases 
do not influence the explicitly CP conserving 
observable.}
\end{center} 
\end{figure}

For the design parameters of the future Linear Collider we choose two
possible levels of performance: a 0.5~TeV machine collecting an
integrated luminosity of 0.5~ab${}^{-1}$ (500~fb${}^{-1}$), and a high
performance version with 1~TeV and 1~ab${}^{-1}$.  Although the
smaller machine might not be able to produce large numbers of higgsino
pairs, the high performance version will extend its mass reach only at
the expense of the gaugino cross sections which drop like $1/s$, as
illustrated in Figure~\ref{fg:cxn}. In Table~\ref{tb:cxn} we present
the set of masses and cross sections for the scenarios (1) and
(4).\smallskip

Typical errors for the mass measurement at a 800~GeV Linear Collider
have been estimated for a scenario very similar to scenario (1),
rendering $0.05/0.07/0.3/0.6$~GeV for the four neutralinos and
$0.035/0.25$~GeV for the charginos~\cite{lincol}. We use these one
sigma error bars for all scenarios, because the typical mass scales
are similar. Moreover, we disregard additional theoretical errors or
higher order corrections; the latter merely shift the theoretical
predictions and affect the currently unknown errors.  The effect of
mismeasured $t$--channel slepton masses on non-polarized total cross
sections is negligible~\cite{sleptonmass}; we assume their exact
determination from the direct production to reduce the number of
observables varied in the analysis. The total cross section for $\nne$
pair production is not part of our set of observables.\bigskip

For the fits, we minimize $\chi^2 = \sum_i (x_{\rm reconstr,i}-x_{\rm
  meas,i})^2/e_{\rm i}^2$ using MINUIT. We choose a Gaussian
probability distribution for the observed values for masses and cross
sections $x_{\rm meas,i}$, disregarding systematic errors. Assuming an
efficiency $\epsilon=10\%$ we obtain $e_{\rm i}^2=\sigma_{\rm
  i}/(\epsilon{\cal L})$ for the cross section measurements. The
result of the inversion is a set of reconstructed theoretical input
parameters $x_{\rm reconstr,i}$. Randomly varying all observables
gives 10000 best fitting sets of reconstructed input parameters,
displayed in each plot. As long as the result is peaked we give the
standard deviation RMS of the pseudo-measurements of the two phases.
\medskip

\begin{figure}[tbh] 
\begin{center}
\includegraphics[width=11.0cm]{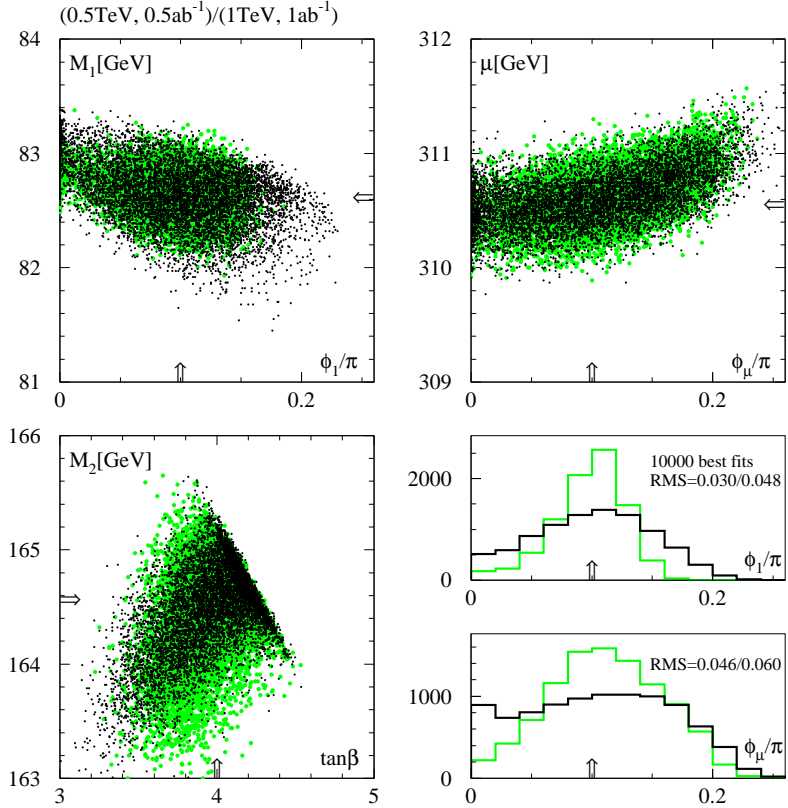} \vspace*{0.0cm}
\caption[]{\label{fg:sugra4} 
Parameters obtained from best fits to masses and cross 
sections with the mSUGRA type parameter choice (1) and 
$\phi_1=0.1 \pi$,
$\phi_\mu=0.1 \pi$. All central values are indicated by
arrows on the axes. The grey (green) and black points 
correspond to the low and high performance collider designs, respectively.
The straight boundary in the $M_2$-$\tan\beta$ plane is an 
artifact from restricting the phases to $[0,\pi]$; this makes 
it easier for MINUIT to find a global minimum in $\chi^2$.}
\end{center} 
\end{figure}

\underline{Modified mSUGRA:} Gaugino mass unification $M_2 \sim 2 M_1$
together with radiative electroweak symmetry breaking $\mu \gg
M_1,M_2$ yields a mass hierarchy in which the heavy higgsinos have
generically smaller cross sections than the light gauginos; their
largest cross section is $\lesssim 200$~fb for $\cpme$ pair
production.  Na\"{\i}vely one might guess that in particular the
neutral higgsinos would have large cross sections, since they couple
to the $s$--channel $Z$ boson, whereas the gauginos are produced
through $t$--channel slepton exchange with the sleptons being heavier
than the $Z$. In fact, the mixing and the phase space factors dominate
the hierarchy of the cross sections.  Thus we have to measure small
cross sections of $\lesssim 70$~fb to determine $\phi_\mu$.  Moreover,
large errors on the higgsino masses do not allow as precise a
determination of $\phi_\mu$ as of $\phi_1$.  Figure~\ref{fg:sugra4}
shows that $\phi_1=0.1 \pi$ can be determined to $\sim 30\%$ with the
0.5~TeV machine, whereas the best fit values for $\phi_\mu=0.1 \pi$
render a RMS value of $46\%$. For the high performance design
the $\phi_\mu$ determination becomes especially difficult. This cannot be
improved by applying cuts on $\chi^2$; all 10000 best fits are of
similar quality.  The $\chi^2$ value in this na\"{\i}ve approach
contains only little information on how closely the reconstructed
values lie to the theoretical input ones.

\begin{figure}[tbh] 
\begin{center}
\includegraphics[width=11.0cm]{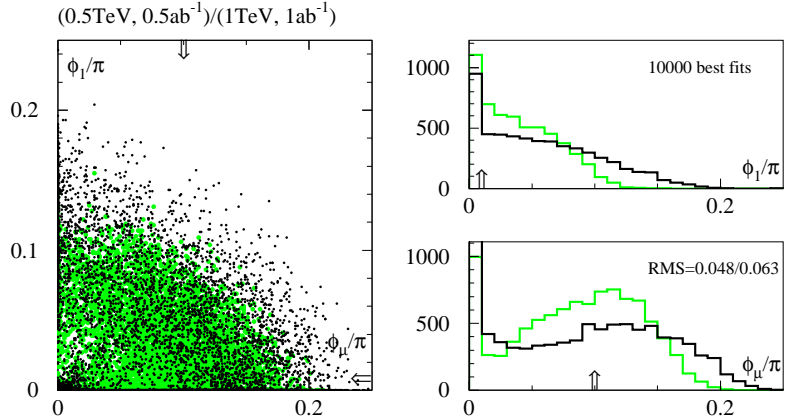} \vspace*{0.0cm}
\caption[]{\label{fg:small4} 
Parameters obtained from best fits to masses and cross 
sections with the mSUGRA type 
parameter choice (2) and $\phi_1=0.01 \pi$,
$\phi_\mu=0.1 \pi$. The central values are indicated by
arrows on the axes. The grey (green) and black points 
correspond to the low and high performance collider designs, respectively.}
\end{center} 
\end{figure}

A very small phase $\phi_1 \lesssim 0.01 \pi$ will be indistinguishable
from zero: the distribution of the fits in Figure~\ref{fg:small4} does
not allow one to extract a non-zero central value of $\phi_1$.
Moreover, the $\phi_\mu$ distribution develops a second peak for zero
phases. This indicates that there are two possible minima in $\chi^2$,
leading to best fits of similar quality. Since this is a drawback of
the weak dependence of masses and cross sections on small phases, the
fits of the mass values in Figure~\ref{fg:sugra4} still vary by only
${\cal O}$(1GeV)\footnote{We have also fitted a zero phase mSUGRA
  scenario $\phi_1=0=\phi_\mu$ leaving the phase values free. The
  obtained RMS values are somewhat dependent on the allowed frame in
  the fit; typical RMS values are 0.03/0.04 for the $\phi_1$
  distribution and 0.06/0.07 for $\phi_\mu$ assuming the low/high
  performance collider design.}. A small value for $\tan \beta$ can be
determined to $\sim 10 \%$ for both of the collider performances.

\begin{figure}[tbh] 
\begin{center}
\includegraphics[width=11.0cm]{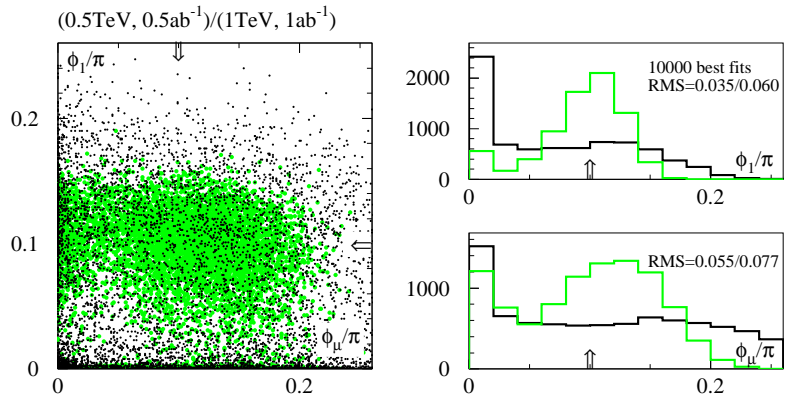} \vspace*{0.0cm}
\caption[]{\label{fg:sugra30} 
Parameters obtained from best fits to masses and cross 
sections at the large $\tan \beta$ mSUGRA 
parameter point (3) with $\phi_1=0.1 \pi$,
$\phi_\mu=0.1 \pi$. All central values are indicated by
arrows on the axes. The grey (green) and black points 
correspond to the low and high performance collider designs, respectively.}
\end{center} 
\end{figure}

The case of $\tan \beta = 30$ in Figure~\ref{fg:sugra30} shows a
similar behavior to Figure~\ref{fg:sugra4}. There is, however, a large
uncertainty in $\tan \beta$; the steep rising of the tangent is not
reflected in the sine and cosine behavior of the mass matrices and
yields a relative error of $\gtrsim 20 \%$ on the measurement of large
$\tan \beta$ values.  This $\tan \beta$ uncertainty influences the
measurement of both of the phases, but $\phi_1 = 0.1 \pi = \phi_\mu$
can still be distinguished from zero and probably be measured at a
0.5~TeV collider.\smallskip

For all mSUGRA type scenarios, the larger number of observed processes
at the high energy collider does not increase the accuracy of the
measurements. As depicted in Figure~\ref{fg:cxn}, gaugino pairs have
large cross sections at the 0.5~TeV machine, and the phase of the
higgsino parameter has a sufficient impact on the $\cpe\cmz$
production cross section and the higgsino masses. Even compensating
for the smaller cross sections by doubling the luminosity of the 1~TeV
collider does not improve the results of our fits. The additional
observables are not sensitive enough to the phases to effectively
contribute to the quality of the fit. However, this only holds as long
as the threshold for $\cpe\cmz$ production is slightly below
0.5~TeV.\smallskip

\begin{figure}[tbh] 
\begin{center}
\includegraphics[width=11.0cm]{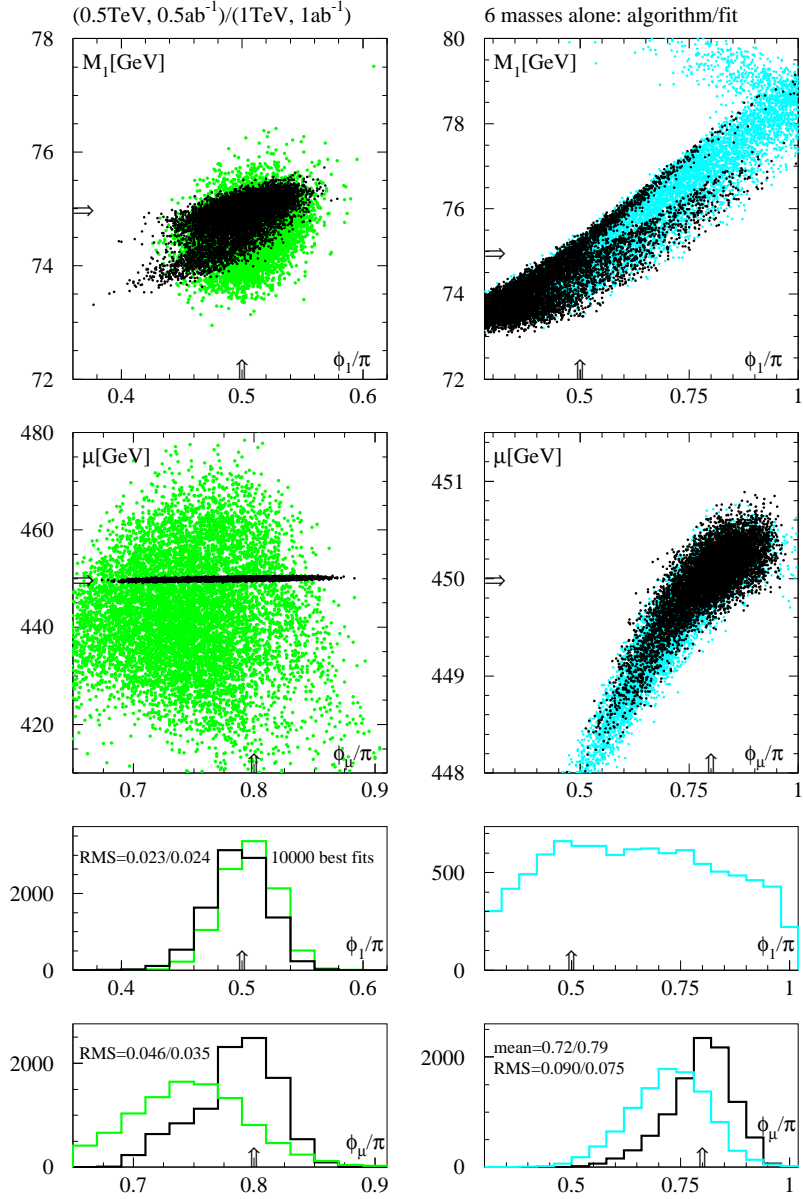} \vspace*{0.0cm}
\caption[]{\label{fg:large} 
Parameters obtained from best fits to masses and cross 
sections (left) and to masses (right) at the large 
phase parameter point (4) $\phi_1=0.5 \pi$,
$\phi_\mu=0.8 \pi$. The central values are again indicated 
by arrows on the axes. The grey (green) and black points 
correspond to the low and high performance collider 
designs, respectively.
For the grey (blue) dots 
in the right column of plots the fit to the masses is
replaced by the algorithm described in the Appendix.}
\end{center} 
\end{figure}

\underline{Large phases:} Avoiding the dipole moment constraints by
fine tuning $\mu$ leads to very heavy higgsinos in the given scenario
(4). They will mainly be produced together with a gaugino. The set of
observable cross sections ranges up to $\sim 170$~fb, even for the
high energy design. Large phases $\phi_1, \phi_\mu$ lead to a
considerable effect on masses as well as on cross sections. The phase
$\phi_1$ can be measured to $\sim 5(7)\%$ at the low (high)
performance machine, see Figure~\ref{fg:large}.  The lower energy
removes the higgsino masses and cross sections from the sample; this
renders the parameter $\mu$ indeterminable. The $\phi_\mu$
distribution peaks around a wrong central value, again showing that
there is no sensitivity to the phase. On the other hand the accuracy
of the $M_1$ and $\phi_1$ measurement is not affected. Using
1~ab${}^{-1}$ at a 1~TeV collider improves the $\phi_\mu$ measurement
to $\lesssim 5\%$.

Since fitting to the large phase scenario yields the best measurements
of the phases we also try to extract the theoretical parameters from
the masses alone. In the right column of Figure~\ref{fg:large} we
compare the fit to the masses to the result of the algorithm described
in the Appendix. Neither the fit to the masses nor the algorithm lead
to a measurement of $\phi_1$: the distribution of values extracted
from the algorithm is entirely flat. The corresponding fit would
prefer arbitrary wrong minima, rendering the result completely
dependent on the setup of the fitting procedure. Both approaches yield
a similar central value and width for $\phi_\mu$ only because the fake
minima for the complete set of parameters give identical $\phi_\mu$
values. This example shows that only a large number of observables can
guarantee a reliable determination of phases, compensating for the
generally weak dependence of the observables on the phase parameters.
\smallskip

\begin{figure}[tbh] 
\begin{center}
\includegraphics[width=11.0cm]{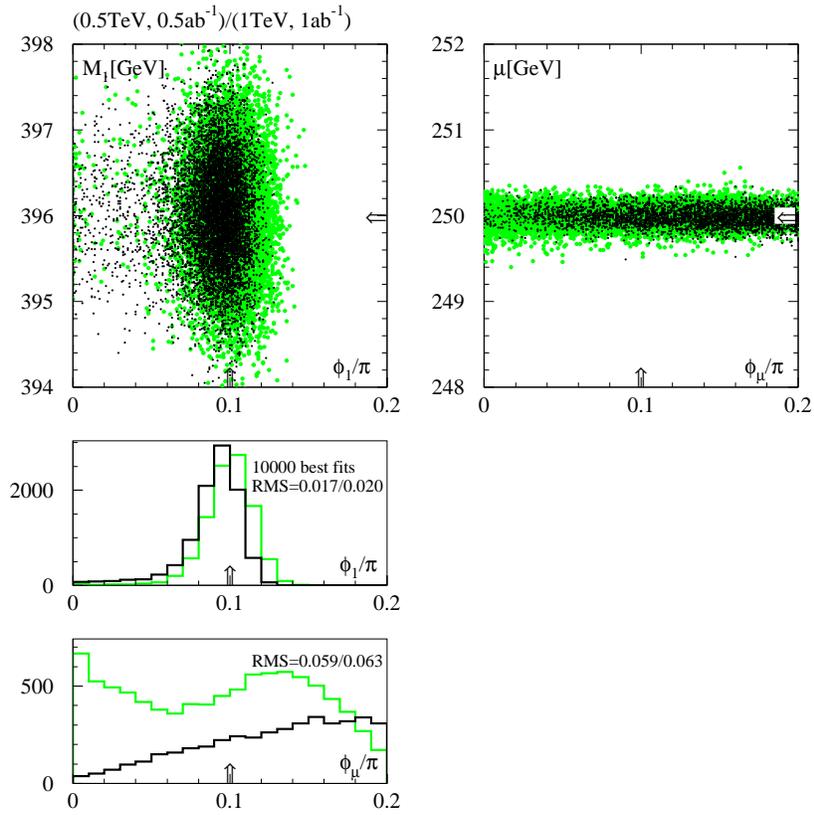} \vspace*{0.0cm}
\caption[]{\label{fg:randall} 
Parameters obtained from best fits to masses and cross 
sections at the wino LSP 
parameter point (5) with $\phi_1=0.1 \pi$,
$\phi_\mu=0.1 \pi$. The central values are indicated by
arrows on the axes. The grey (green) and black points 
correspond to the low and high performance collider 
designs, respectively.}
\end{center} 
\end{figure}

\underline{Modified wino LSP:} Potentially dangerous features of the
wino LSP scenario are large values for $M_1$ and $\mu$, the complex
entries in the mass matrix. The phases could effectively decouple.
This, however, merely affects the accuracy of the fitted $M_1$ values
displayed in Figure~\ref{fg:randall}. It is limited by the large
errors of the heavy neutralino/chargino mass measurements. A small
absolute value of $\mu$ on the contrary can easily be determined from
the masses alone. Due to strong mixing in the mass matrices, $\phi_1$
can be measured with $\lesssim 20$ percent uncertainty, whereas
fitting the phase $\phi_\mu$ relies on the small mixed cross sections
and might only distinguish a finite value from zero.

\section{Summary}\label{sec:three}  

For several different scenarios we estimate a possible straightforward
measurement of CP violating supersymmetric phases. Small phases are
added to mSUGRA models and to a wino LSP model: typical values around
$\pi/10$ for $\phi_1,\phi_\mu$ can be determined from total cross
sections and masses by simple fits; they will not be hidden by
anticipated statistical errors. However, very small phases, of the
order of $\pi/100$, as preferred by electric dipole moment
analyses~\cite{small,toby}, can hardly be distinguished from zero. In
this case, the sample of observables behaves like a CP conserving set,
and the mass parameters and $\tan \beta$ are accessible to a high
degree of accuracy~\cite{old}. Large phases can easily be determined
if the corresponding gaugino/higgsino states are produced. For the
model under consideration~\cite{kane} the set of masses and cross
sections at a 1~TeV collider can be used to determine the values to
a few percent. However, the set of masses alone is heavily affected by
experimental errors. There still is a remaining sign ambiguity of the
phases.  Forward-backward asymmetries, left-right asymmetries, and
other explicitly CP non-conserving observables will considerably
improve our estimates~\cite{old,zerwas}.  A more sophisticated
analysis including some detector simulation would have to be carried
out to determine where the lower limits for detectable very small
phases will finally lie.


\acknowledgements

T.P. thanks D.~Zerwas, P.M.~Zerwas and in particular T.~Falk
and G.~Blair for very helpful discussions and comments on this
paper. T.L. thanks Zhou Mian-Lai for discussions. This research was
supported in part by the University of Wisconsin Research Committee
with funds granted by the Wisconsin Alumni Research Foundation and in
part by the U.~S.~Department of Energy under Contract
No.~DE-FG02-95ER40896.


\newpage
\appendix 
\section{Inversion Algorithm}

From the complete set of neutralino/chargino masses we can
unambiguously determine the underlying mass parameters.\footnote{Ways
  have been shown to extract these parameter from a smaller set of
  masses~\cite{kneur}. However, the complete set is well suited to
  illustrate the result from the fits.} In the following we denote 
the absolute values $|M_1|, |\mu|$ by $M_1,\mu$. 
First we calculate all parameters
as a function of $M_2$:
\begin{alignat*}{9}  
\mu =& \left[ \sum_{i=1}^2 M_{\chi_i^{\pm}}^2 -M_2^2 -2 M_W^2
       \right]^{1/2} \qquad \qquad \qquad &
 M_1 =& \left[ \sum_{i=1}^4  m_{\tilde \chi^{0}_i}^2 -M_2^2 
           - 2\mu^2 -2 M_Z^2 \right]^{1/2}  \\
\sin 2 \beta =& \left[ \frac{-Y - \sqrt{Y^2-4 X Z}}{2X} 
                \right]^{1/2} & \\
\cos\phi_1 =& C_2 + F_2 \sin^22\beta &
\cos\phi_\mu =& \frac{C_0 +F_0 \sin^22\beta}{\sin2\beta}
\end{alignat*}
using 
\begin{alignat*}{9}  
X=& \; 2 s_w^2 c_w^2 M_1 M_2 M_Z^4 F_2 \\
Y=& \; 2 s_w^2 c_w^2 M_1 M_2 M_Z^4 C_2
 - 2 s_w^2  M_1 M_2^2 \mu M_Z^2 F_1
 - 2 c_w^2  M_1^2 M_2 \mu M_Z^2 F_0
 + c_w^4 M_1^2 M_Z^4 
 + s_w^4 M_2^2 M_Z^4 \\
Z=& \; - 2 s_w^2  M_1 M_2^2 \mu M_Z^2 C_1
       - 2 c_w^2  M_1^2 M_2 \mu M_Z^2 C_0
       + M_1^2 M_2^2 \mu^2 -
         \frac{\Pi_{i=1}^4 m_{\tilde \chi^{0}_i}^2}{\mu^2} \\
C_0 =& \; \frac{1}{8 M_W^2 M_2 \mu}
  \left[ \left(M_{\chi_2^{\pm}}^2 -  M_{\chi_1^{\pm}}^2 \right)^2
  - \left( M_2^2-\mu^2 \right)^2 
  - 4 M_W^2 \left( M_2^2 +\mu^2 + M_W^2 \right) \right] \\
C_1 =& \; \frac{1}{4 s_w^2 M_1 \mu M_Z^2}
  \left[ \sum_{i=1}^4  m_{\tilde \chi^{0}_i}^4 -2 M_Z^4
  - M_1^4-M_2^4-2\mu^4 
  - 4 M_Z^2 \left( s_w^2 M_1^2  +  \mu^2 
  + c_w^2 M_2^2 + c_w^2 M_2 \mu C_0 \right) \right] \\
C_2 =& \; \frac{1}{6s_w^2 c_w^2 M_1 M_2 M_Z^4}
  \left[ \sum_{i=1}^4  m_{\tilde \chi^{0}_i}^6 - 6 \mu^2 M_Z^4
  - M_1^6-M_2^6- 2\mu^6 -2 M_Z^6 
  \right. \\ & \left. \phantom{\frac{1}{6s_w^2 c_w^2 M_1 M_2 M_Z^4}}
  - 3 s_w^2 ( 2 + s_w^2 ) M_1^2 M_Z^4
  - 3 c_w^2 ( 2 + c_w^2 ) M_2^2 M_Z^4
  \right. \\ & \left. \phantom{\frac{1}{6s_w^2 c_w^2 M_1 M_2 M_Z^4}}
  - 6 c_w^2 M_2 \mu M_Z^2 \left( M_2^2+\mu^2+ 2 M_Z^2 \right) C_0 
  - 6 s_w^2 M_1 \mu M_Z^2 \left( M_1^2+\mu^2+2 M_Z^2 \right) C_1 
  \right. \\ & \left. \phantom{\frac{1}{6s_w^2 c_w^2 M_1 M_2 M_Z^4}}
  - 6\mu^4 M_Z^2
  - 6 s_w^2 M_1^2 M_Z^2 \left( M_1^2 +\mu^2 \right)
  - 6 c_w^2 M_2^2 M_Z^2 \left( M_2^2 +\mu^2 \right) \right] \\
F_2 =& \; -\frac{\mu^2}{ 2 s_w^2 c_w^2 M_1 M_2}
          -\frac{\mu (M_1^2+\mu^2+ 2 M_Z^2) F_1}{ c_w^2 M_2 M_Z^2}
          -\frac{\mu  (M_2^2+\mu^2+ 2 M_Z^2 ) F_0}{ s_w^2 M_1 M_Z^2} 
\end{alignat*}
Here $F_0=M_W^2/(2 M_2 \mu)$, $F_1=M_2 c_w^2 F_0/(M_1 s_w^2)$, and
$s_w,c_w$ stand for the sine/cosine of the weak mixing angle.
After computing these parameters, 
\begin{equation*}
\cos(\phi_1+\phi_\mu)= \frac{C_1+F_1\sin^22\beta}{\sin2\beta}
\end{equation*}
serves as a self consistency relation and
thereby determines $M_2$.


\bibliographystyle{plain}

\begin{table}[tbh] 
\begin{center}
\begin{tabular}{c|r|r||c|r|r||c|r|r||c|r|r}
 $m_{\tilde{\chi}}$~[GeV] & (1)  &  (4) &
 $\sigma_{\rm tot}$~[fb]  & (1)  &  (4) &
 $\sigma_{\rm tot}$~[fb]  & (1)  &  (4) &
 $\sigma_{\rm tot}$~[fb]  & (1)  &  (4)  \\[1mm] \hline
 $\mne$        &  77.8  &   74.6  & 
 ($\nne \nne$) &  98.5  &   91.6  &
 $\nne \nnd$   &   3.3  &    0.2  &
 $\nnd \nnd$   &  0.01  &   10$^{-5}$  \\
 $\mnz$        & 143.0  &   96.3  &
 $\nne \nnz$   &  32.0  &   70.8  &
 $\nne \nnv$   &   5.7  &    3.2  &
 $\nnd \nnv$   &  34.4  &   27.3  \\
 $\mnd$        & 315.6  &  450.2  &
 $\nnz \nnz$   &  66.3  &   30.5  &
 $\nnz \nnd$   &   8.6  &    1.0  &
 $\nnv \nnv$   &   0.5  &  0.002  \\
 $\mnv$        & 342.6  &  465.8  &
               &        &         &
 $\nnz \nnv$   &  12.7  &    2.1  &
               &        &         \\
 $\mce$        & 141.3  &   94.2  &
 $\cpe \cme$   & 142.5  &  170.3  &
 $\cpe \cmz$   &  21.6  &    5.2  &
 $\cpz \cmz$   &  87.1  &   59.6  \\
 $\mcz$        & 341.3  &  462.4  &
               &        &         &
               &        &         &
               &        &         \\
\end{tabular} \vspace*{0.3cm}
\caption[]{\label{tb:cxn}
Neutralino/chargino masses and cross sections 
at a 1~TeV Linear Collider
for the modified mSUGRA scenario (1) and the large phase 
model (4). The large $\nne \nne$ cross 
section is not used in the analysis.}
\end{center}
\end{table}

\end{document}